\documentclass[conference, a4paper]{IEEEtran}
\IEEEoverridecommandlockouts
\usepackage{cite}
\usepackage{amsmath,amssymb,amsfonts}
\usepackage{algorithmic}
\usepackage{graphicx}
\usepackage{textcomp}
\usepackage{xcolor}
\usepackage{etoolbox}

\makeatletter
\patchcmd{\@makecaption}
  {\scshape}
  {}
  {}
  {}
\makeatother

\begin{document}
\title{Deep Neural Networks based Invisible Steganography for Audio-into-Image Algorithm}


\author{
\IEEEauthorblockN{Quang Pham Huu\IEEEauthorrefmark{1}, Thoi Hoang Dinh\IEEEauthorrefmark{2}, Ngoc N Tran\IEEEauthorrefmark{1}\IEEEauthorrefmark{3}, Toan Pham Van\IEEEauthorrefmark{1}, Thanh Ta Minh\IEEEauthorrefmark{1}\IEEEauthorrefmark{4}}
\IEEEauthorblockA{\IEEEauthorrefmark{1}R\&D Lab, Sun* Inc\\
\{pham.huu.quang, tran.ngo.quang.ngoc, pham.van.toan, ta.minh.thanh\}@sun-asterisk.com\\
\IEEEauthorrefmark{2} R\&D Center, Samsung SDS Vietnam\\
hd.thoi@samsung.com\\
\IEEEauthorrefmark{3} Rensselaer Polytechnic Institute, USA\\
trann2@rpi.edu\\
\IEEEauthorrefmark{4} Le Quy Don Technical University, 236 Hoang Quoc Viet, Bac Tu Liem, Ha Noi\\
thanhtm@mta.edu.vn}
}

\maketitle

\begin{abstract}
In the last few years, steganography has attracted increasing attention from a large number of researchers since its applications are expanding further than just the field of information security. The most traditional method is based on digital signal processing (DSP), such as least significant bit (LSB) encoding. Recently, there have been some new approaches employing deep learning to address the problem of steganography. However, most of the existing approaches are designed for image-in-image steganography. In this paper, the use of deep learning techniques to hide secret audio into the digital images is proposed. We employ a joint deep neural network architecture consisting of two sub-models: the first network hides the secret audio into an image, and the second one is responsible for decoding the image to obtain the original audio. Extensive experiments are conducted with a set of 24$K$ images and the VIVOS Corpus audio dataset\footnote{https://ailab.hcmus.edu.vn/downloads}. Through experimental results, it can be seen that our method is more effective than traditional approaches. The integrity of both image and audio is well preserved, while the maximum length of the hidden audio is significantly improved.

\end{abstract}

\vspace{-10pt}
\section{Introduction}
In modern era, computers and internet play an important role in many aspects of life, especially in the field of information technology. Along with the proliferation of the internet, there have been growing concerns with regards to information security. There are two main solutions to this problem: cryptography and steganography. Unlike cryptography which aims to protect the content of the message, steganography involves concealing secret information inside common forms of data.
It is generally used to transmit confidential information to people who already are aware of its existence, while others still see it as normal data.

Today, on the internet, billions of photos and audio files are transmitted every day. This fact makes steganography more and more popular and widely used. In addition to security, it is also used in the entertainment and software industries as a watermarking technique on images, music or digital software for copyright protection, impersonation detection, duplication prevention, content validation; allowing monitoring or tracing of illegal copies as well as monitoring advertising.

In this paper, we address the topic of audio-into-image steganography which aims to hide secret audio in digital image files. Unlike hiding image in image or hiding audio in audio, our work is more difficult because of the fact that audio and images are of different domains. Audio data are one-dimensional matrix representing a series of amplitude in time domain, often in PCM-16 bit format, and its values range from $-2^{15}$ to $2^{15}-1$. On the other hand, image is a three-dimensional matrix representing light intensity, and its values range from 0 to 255 (8-bit image). One of the most common approaches is to directly encode the secret audio in least significant bits of each pixel in the image. However, this method has been proved to be ineffective when the number of bits needed to hide is larger than the number of bits that can be changed in the cover data. If more bits are modified, the integrity of the the cover data is no longer preserved. That is the main obstacle in hiding audio data inside image.

In light of these problems, we propose a Deep Convolutional Neural Network (DCNN) model which is capable of both hiding audio into image and revealing the original audio. The workflow of said model is demonstrated in Figure~\ref{workflow}. The secret audio is pre-processed and concealed in the cover image to generate the container image. The container image is, then, used to retrieve the original secret audio.
With our approach, the difference between pairs of cover-container image and secret-revealed audio is practically unnoticeable.

To sum up, our main contributions are: (1) Proposing the use of DNN to address the problem of hiding secret audio into digital images. (2) Comparing different audio data pre-processing methods in term of audio and image differences. (3) Showing that our method can hide longer audio than traditional methods, and that the difference between the cover image and the container image cannot be easily recognized by human's eyes.

The rest of the paper is organized as follows. Section 2 presents a brief review of related work. In section 3, the data pre-processing method and proposed model architecture will be discussed. The data preparation for the experiment and system setup are mentioned in Section 4. Experimental results and evaluation are presented in Section 5. Finally, we conclude the paper in Section 6.

\section{Related work}
\subsection{Traditional steganography techniques}
Steganography has a quite long history of development. Thorough surveys of steganography's techniques and applications can be found in \cite{CHEDDAD2010727, li2011survey}. Methods in ancient steganography are pretty basic: the secret message is physically hidden inside another data form. These methods are simple, and thus, they can be easily detected. As a result, more secure steganography techniques are desired. Thanks to advances in digital signal processing, many digital steganography techniques have been developed. The earliest method is proposed by Kurak and McHugh \cite{kurak1992cautionary} who managed to embed data into 4 least significant bits (LSB) of an image. Hiding secret message inside other data forms other than image is possible, too. In \cite{hernandez2006steganography},  C. Hernandez, \textit{et. al} succeeded in hiding data in HTML and XML documents as well as executable file (\texttt{.exe}).
Or in \cite{hosmer2006discovering}, Chet Hosmer also employ the LSB technique to hide data in GIF and JPEG format image or even music file as well.

\subsection{Deep learning in steganography}
Though LSB-based steganography is widely used, it has some major drawbacks. The main disadvantage of the LSB-based mechanism is that it lacks robustness when facing steganalysis. In order to address the existing problem of LSB technique, approaches employing deep neural networks have been proposed. One of the earliest works is Imran Khan's, where feature representations can be automatically learned by a network containing several convolutional layers \cite{khan2010neural}. In 2017, Shumeet Baluja used a neural network architecture consisting of three convolutional networks to hide secret image in public image of the same size \cite{NIPS2017_6802}. Recently, Jiren Zhu \textit{et. al} proposed an end-to-end deep neural network for both encoding a hidden message as bit string inside an image and decoding the encoded image to retrieve the original message \cite{zhu2018hidden}. Jiren Zhu also proposed the adversarial network alongside the encoder-decoder pair. The adversarial network is able to predict whether a given image contains some secret message and can help improve the quality of the encoded image.
    
\subsection{Audio-into-image steganography}
Audio steganography is a brand of the wider steganography field which aims to covertly hide secret message inside audio files. Much effort has been devoted to address this problem. In literature, some typical works can be found in \cite{cvejic2002wavelet,gopalan2003audio,cvejic2004increasing}.
As for hiding audio file itself in other data forms, the common idea is to apply Short Time Fourier Transform (STFT) to get the spectrogram of the secret and carrier audio and use a pair of deep neural networks for both encoding and decoding. In \cite{ye2019heard}, Denpan Ye \textit{et. al} use a generative adversarial network-based architecture which contains three parts: encoder, decoder, and steganalyzer. Denpan Ye's proposed technique can not only encode and decode audio data but also classify whether an audio file contains some secret message or not. In term of audio-into-image steganography, there have been only a few attempts since audio and image data are of different domains. Most recent research can be found in \cite{santosa2005audio, kaul2013audio}. The two mentioned works both use wavelet transform to compress speech data and LSB technique as embedding algorithm. These methods were proposed quite a long time ago and not much progress has been seen since then. The lack of method for effective audio-into-image steganography is a huge motivation for us. STFT is used as audio processing technique whereas deep convolutional neural network is employed in encoding and decoding phase. Combining the two, our method can benefit from the learning capacity of deep networks and the information extraction of digital signal processing technique.

\begin{figure}
    \centering
    \includegraphics[width=0.45\textwidth,height=4.5cm]{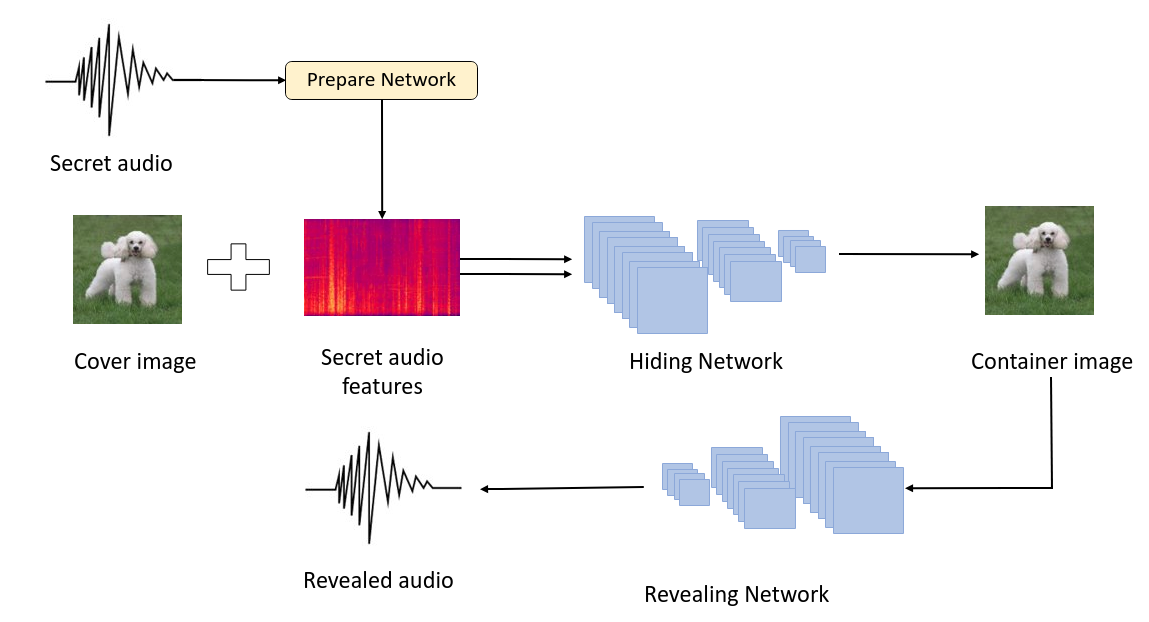}
    \caption{\label{fig:my-label} The general architecture flow.}
    \label{workflow}
\end{figure}

\section{Proposed method}
\subsection{Data Preprocessing}

Normalizing images is quite simple; the value of each pixel is divided by 255 to obtain a new matrix in the value domain from 0 to 1. However, for audio data, there is slightly different processing technique. Though amplitude of audio data in 16-bit PCM format ranges from $-2^{15}$ to $2^{15}-1$. That makes traditional techniques such as min-max scaling, mean-standard deviation normalization not effective. As such, multiple experiments which focus on analyzing histogram of audio are carried out to find the suitable normalization method.

Two methods of audio data pre-processing are conducted.
\begin{itemize}
 \item Method-1: Raw audio data is normalized and reshaped to a new 3D matrix of the same shape as three color channels image.
 \item Method-2: Short-time Fourier transform (STFT) is applied to transform audio to the frequency domain. STFT is a sequence of Fourier transforms of a windowed signal. STFT provides the time-localized frequency information for situations in which frequency components of a signal vary over time. The transformed data is now a 3D matrix with two channel instead of three channels as the first method \cite{Griffin1984SignalEF, Nawab:1987:SFT:42739.42745}.
\end{itemize}

\subsection{Model architecture}
As mentioned in the previous sections, we build a DNN model, specifically, a convolutional neural network (CNN) which are capable of hiding audio into image and revealing the original audio.
There are two sub-models, the first model consists of the prepare network and hiding network; the second one is the reveal network:
\begin{itemize}
\item Encoding submodel:
\begin{itemize}
\item Prepare network: appropriately extracts the characteristics of the audio before it is concatenated with the image.
 \item Hiding network: hides the secret audio features into the cover image to generate the container image.
\end{itemize}
 \item Decoding submodel:
 \begin{itemize}
 \item Reveal network: decodes the container image to obtain the original audio.
 \end{itemize}
\end{itemize}
Our overall architecture is similar to Baluja \textit{et. al}'s method\cite{NIPS2017_6802}. The general architecture is demonstrated in Figure~\ref{workflow}.
However, each component of our proposed convolutional neural network is modified to be more suitable for audio data and to help to reduce training time. The difference in the audio pre-processing techniques also leads to some changes in the overall architecture. Some modifications are required to make the overall architecture more suitable for each pre-processing method. All three network components use the same architecture which we named the base model. This base model is used for both methods of data pre-processing.
\subsubsection{Base model}
\begin{figure}
    \centering
    \includegraphics[width=0.45\textwidth,height=8.5cm]{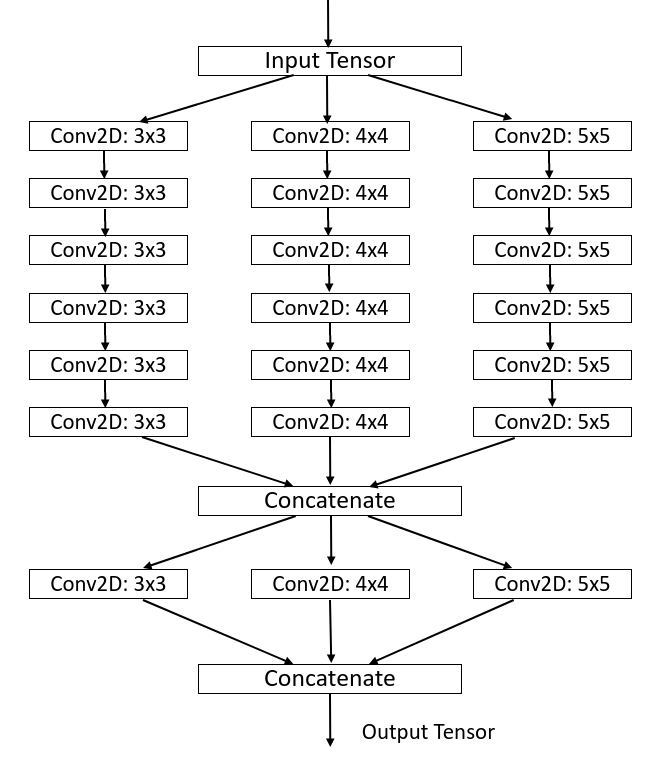}
    \caption{\label{fig:my-label} The base model architecture (3 parallel blocks of 5 convolutions stacked together with kernel sizes of $3$, $4$, $5$).}
    \label{ablock}
\end{figure}

In order for the network to learn many feature levels, the convolution stacked blocks are built. A block is an architecture of  5 convolutions stacked together with the same kernel size. There are three parallel blocks with kernel sizes of 3$\times$3, 4$\times$4, and 5$\times$5 that make network architecture learn more diverse features. The output of three block are concatenated and passed through a similar architecture but shallower, with only one convolution layer. Finally, the output are concatenated and goes through another convolution layer to get the expected output size. The base model architecture is demonstrated in Figure~\ref{ablock}.

\subsubsection{Prepare network}
The prepare network configuration is simply the base model. It is responsible for extracting the characteristics of audio data more appropriately before it is concatenated with the image. When using the data pre-processing method 1, the prepare network input is a 3D tensor of shape 255$\times$255$\times$3. For the data pre-processing method 2, the network input is a 3D tensor of shape 255$\times$255$\times$2. In both cases, the expected output is tensor of shape 255$\times$255$\times[\textit{feature\_maps}]$. After the audio features are extracted by the prepare network, it is concatenated to the cover image to become the input for the hiding network.
\subsubsection{Hiding network}
Similar to the prepare network, the hiding network also has the same architecture as the base model. With input being a tensor containing information of both cover image and secret audio, the network is responsible for hiding the secret audio features into the cover image to generate the container image. Network weights are trained so that the generated container image looks the same as the cover image.
\subsubsection{Reveal network}
The reveal network is designed to be capable of decoding the container image and obtaining the original audio.
As for the reveal network, we specifically use linear activation function in the last layer instead of rectified linear unit (ReLU) due to the characteristics of audio data. For each pre-processing method, the output of the network is designed to match the decoding process. The expected output is a 3D tensor of shape 255$\times$255$\times$3 with raw audio, 3D tensor of 255$\times$255$\times$2 with STFT pre-process. With the STFT method, we used Inverse Short Time Fourier transform algorithm (ISTFT) \cite{Griffin1984SignalEF} to recreate the original sound.

\subsection{Error measurement}
As mentioned above, our main purpose is to build a model which is capable of both hiding audio into image and recovering the secret audio. The goal of hiding network is to hide secret audio $S$ in the cover image $C$. After hiding $S$, the newly created image contains information about secret audio, called the container image or the hidden image $H$. The container image is passed onto the reveal network to obtain the original audio, called the revealed audio $O$. To measure the differences between the cover image and the container image and between the secret audio and revealed audio, we use mean square error (MSE) as metric.
The loss function used is a weighted sum of the mean squared errors of the hiding network and the reveal network.

\begin{equation}
L(C, H, S, O) = \frac{\alpha \mathrm{MSE}(C, H) + \beta \mathrm{MSE}(S, O)}{\alpha + \beta}
\end{equation}
where $MSE(C, H)$ is the mean square error between cover image $C$ and hidden image $H$, $MSE(S, O)$ is the mean square error between secret audio $S$ and revealed audio $O$. $\alpha$ and $\beta$ are mathematical constant to modify the trade-off between optimizing audio loss and image loss. When the value of $\alpha$ is bigger, the high quality of the container image can be achieved. On the other hand, when the value of $\beta$ is bigger, the quality of the extracted audio is better. During the training phase, parameters of the network is updated so that $L(C, H, S, O)$ is minimized.
\vspace{-5pt}

\section{Dataset and system setup}
\subsection{Dataset}
We conduct experiments on two datasets: Vivos dataset and a popular set of image data described below.
Vivos is a public dataset that includes 12,420 audio of Vietnamese voices with the sample rate of 16kHz. Duration of the audio ranges from 1$-$18 seconds. The image dataset consists of a large number of photos from Kaggle\footnote{https://www.kaggle.com}
competitions including the Flower, Fruits, Dogs and Cats, and Stanford Dogs datasets. In total, the collected image dataset contains up to 24,000 images. 80\%, 10\%, 10\% of the dataset are used in the training, validating, and testing phase, respectively.
\subsection{System setup}
Our experiment is conducted on a computer with Intel
Core i5-7500 CPU @3.4GHz, 16GB of RAM, GPU GeForce
GTX 1080 Ti. The proposed architecture is implemented with the Keras\footnote{https://keras.io/} framework.

\section{Results and evaluation}
\subsection{Evaluation metrics}
To evaluate the integrity of the cover image before and after secret audio is hidden, we used the sum of squares error (SSE) metric on 1000 images. That means we calculate MSE per pixel, per channel on 1000 pairs of images and sum them up.
\begin{equation}
\mathrm{SSE} = \sum_{1}^{1000} \mathrm{MSE}_\textit{per\_pixel,\ per\_channel}
\end{equation}
Suppose that the height and width of image are $H$ and $W$, the MSE per pixel, per channel is defined as:
\begin{equation}
\mathrm{MSE}_\textit{per\_pixel,\ per\_channel} = \frac{1}{3\times H \times W}\sum(y-\hat{y})^{2}, 
\end{equation}
where $y$, $\hat{y}$ is the value of each pixel in each color channel in the cover and container image, respectively. For audio, we use the Pearson correlation coefficient (PCC) to evaluate the integrity of information after the process of hiding and revealing. The Pearson correlation coefficient measures the linear relationship between two data distribution. The correlation coefficient is calculated as follows:

\begin{equation}
\mathrm{r} = \frac{\sum(x - m_x)(y - m_y)}{\sqrt{\sum(x-m_x)^2 \sum(y-m_y)^2}}
\end{equation}
where $m_x$ is the mean of the vector $x$ and $m_y$ is the mean of the vector $y$.

The value of correlations value range from -1 and +1 with 0 implying no correlation. Correlations of -1 or +1 imply an exact linear relationship. Two audio is considered to be completely the same when the correlation coefficient is 1 or 100\% \cite{10.1093/biomet/6.2-3.302,10.2307/2346598}.

\subsection{Evaluation result}
\begin{figure}
    \centering
    \includegraphics[width=0.45\textwidth,height=4cm]{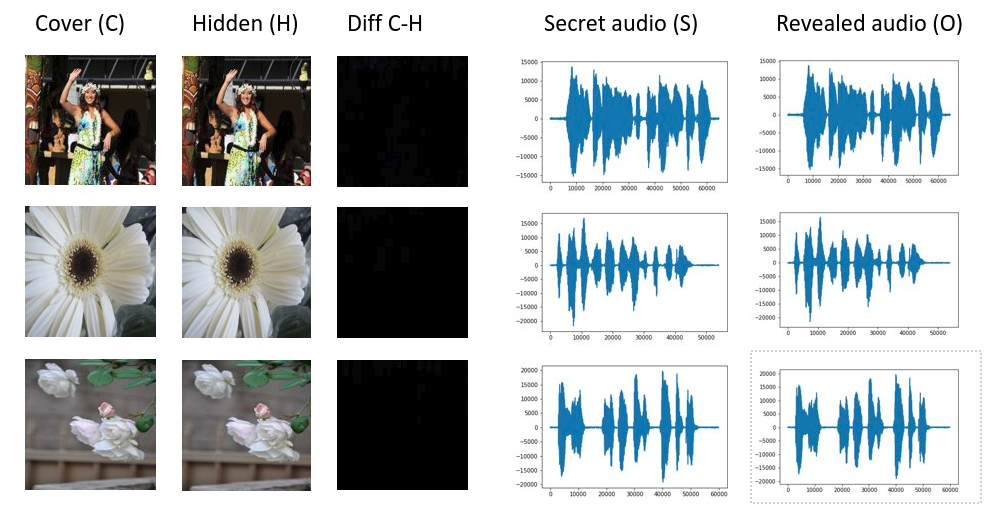}
    \caption{Results using STFT pre-processing}
    \label{stft_result}
\end{figure}
\begin{figure}
    \centering
    \includegraphics[width=0.45\textwidth,height=4cm]{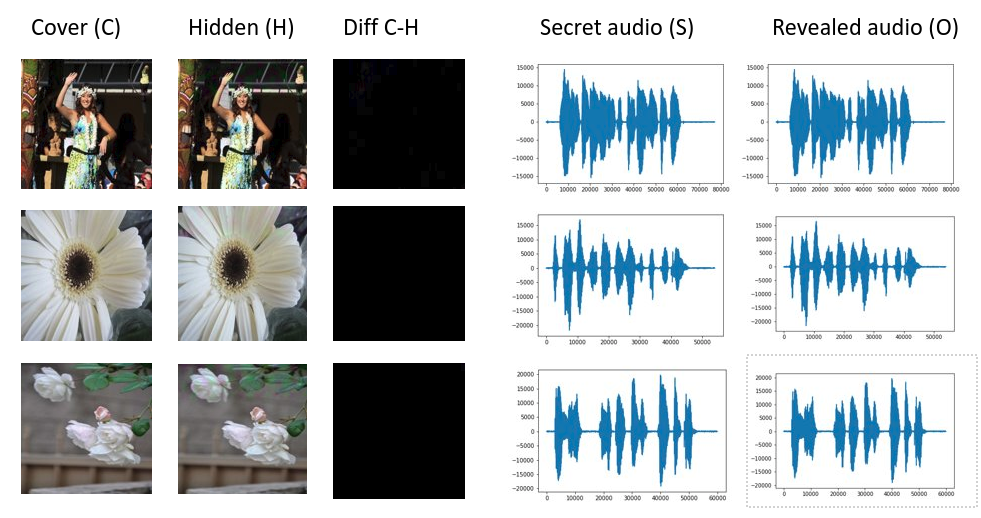}
    \caption{Results using raw data pre-processing}
    \label{raw_result}
\end{figure}

Experimental results are presented in Tables~\ref{tmp}. It shows that our methods could be suitable for audio-into-image steganography.
The average correlations across different parameter settings are all above 99\%, and can be as high as 99.93\%, while not trading off any perceivable image quality using STFT pre-processing. Across the board, with identical parameter settings, audio correlation metrics are slightly better with STFT pre-processing comparing to the raw audio method, while the SSE metrics are much better.
Moreover, we can compare our methods with least significant bits (LSB) method.  
LSB replaces some of LSB of the cover image with data from the secret audio. The drawbacks of LSB is that the number of bits that can be hidden depends on the number of LSB bits that could change for each pixel; while with our method, the network can also hide information in the correlations between pixels, resulting in both a better hiding capacity and a more natural container image.
Specifically, with our DFT pre-processing method, a 4-second-long audio can be hidden into a 255$\times$255 image. With the raw audio data, the length of audio data can be up to 12 seconds.
Taking both hiding-recovering quality and source audio length into consideration, we can see that STFT pre-processing is a clear-cut winner. As a bonus, it also provides us a way to visualize the audio quality loss with a spectrogram XOR figure.

The Figure~\ref{stft_result} and Figure~\ref{raw_result} show that some results of our method using STFT pre-processing and raw data pre-processing methods.
\begin{table}[t]
\centering
\caption{SSE (smaller is better) and correlation coefficient (greater is better) with raw and STFT pre-processing techniques}
\def\tablename{table}
\begin{tabular}{|c|c|c|c|}
\hline
Pre-process & \textbf{$\alpha/\beta$} & SSE & Average of correlation\\ \hline
& 15 / 1 & 0.3744 & 99.83\% \\
\cline{2-4}
&10 / 1 & 2.0379 & 99.30\% \\
\cline{2-4}
Method-1: Raw&1 / 1 & 2.4072 & 99.84\% \\
\cline{2-4}
&1 / 2  & 10.7196 & 99.74\% \\
\cline{2-4}
&1 / 10  & 14.8777 & 99.90\% \\
\hline
&10 / 1 & 0.0192 & 99.43\% \\
\cline{2-4}
&2 / 1  & 0.0135 & 99.91\% \\
\cline{2-4}
Method-2: STFT &1 / 1  & 0.0334 & 99.85\% \\
\cline{2-4}
&1 / 2  & 0.0539 & 99.86\% \\
\cline{2-4}
&1 / 10  & 0.1393 & 99.93\% \\ \hline
\end{tabular}
\label{tmp}
\end{table}

\section{Conclusion}
In this paper, the use of deep learning technique in audio-into-image steganography is explored. Two convolutional neural network are designed to work in pairs to hide secret audio into public image and retrieve the original audio data from the encoded image. Also, two different audio processing methods and some neural network architecture modifications are introduced to make current approach more suitable for audio data. Extensive experiments with different settings are carried out to support the superiority of our proposed method. Through the experimental result, it has been verified that the integrity of both original image and audio is preserved. The length of hidden audio is also improved when compared to traditional steganography techniques. As such, the proposed method is well capable of addressing the problem of audio-into-image steganography.

\bibliographystyle{plain}
\bibliography{ref}
\end{document}